\documentclass{article}

\usepackage{arxiv}

\usepackage[utf8]{inputenc} % allow utf-8 input
\usepackage[T1]{fontenc}    % use 8-bit T1 fonts
\usepackage{hyperref}       % hyperlinks
\usepackage{url}            % simple URL typesetting
\usepackage{graphicx}
\usepackage{booktabs}       % professional-quality tables
\usepackage{amsfonts}       % blackboard math symbols
\usepackage{nicefrac}       % compact symbols for 1/2, etc.
\usepackage{microtype}      % microtypography
\usepackage{lipsum}

\title{Triplet loss based embeddings for forensic speaker identification in Spanish}
\date{01 december}

\author{
  Emmanuel ~Maqueda \\
  Facultad de Estudios Superiores Cuautitlán\\
  Universidad Nacional Autónoma de México\\
  \texttt{emmaqueda@comunidad.unam.mx} \\
   \And
 Javier ~Alvarez-Jimenez\\
  Universidad Abierta y a Distancia de México\\
  \texttt{javier.alvarezjim@nube.unadmexico.mx} \\
   \AND
   Carlos ~Mena \\
  University of Malta\\
   \texttt{carlos.hernandez@um.edu.mt} \\
   \And
  Ivan ~Meza* \\
  Instituto de Investigaciones \\ en Matemáticas Aplicadas y en Sistemas\\
  Universidad Nacional Autónoma de México\\
  \texttt{ivanvladimir@turing.iimas.unam.mx} \\
}

\begin{document}
\maketitle

\begin{abstract}
With the advent of digital technology, it is more common that committed crimes or legal disputes involve some form of speech recording where the identity of a speaker is questioned~\cite{rose1997identifying}. In face of this situation, the field of forensic speaker identification has been looking to shed light on the problem by quantifying how much a speech recording belongs to a particular person in relation to a population. In this work, we explore the use of speech embeddings obtained by training a CNN using the triplet loss. In particular, we focus on the Spanish language which has not been extensively studies. We propose extracting the embeddings from speech spectrograms samples, then explore several configurations of such spectrograms, and finally, quantify the embeddings quality. We also show some limitations of our data setting which is predominantly composed by male speakers. At the end, we propose two approaches to calculate the Likelihood Radio given out speech embeddings and we show that triplet loss is a good alternative to create speech embeddings for forensic speaker identification.
\end{abstract}

% keywords can be removed

\section{Introduction}

The use of Triplet loss~\cite{schroff2015facenet} was popularized with the introduction of the FaceNet architecture which was aimed to face identification tasks. This loss allows to train a neural network, commonly a Convolutional Neural Network (CNN), to produce a vector representation of an image. The goal is that the neural network learns a mapping from face images to an euclidean space, it is desirable that distances among face image positions directly correspond to a measure of face similarity. Within this setting a desired outcome is that images of the same face will cluster together and images of different faces will be separated by a margin. Triplet loss has been applied into different types of images: objects~\cite{wang2014learning}, person re-identification~\cite{cheng2016person}, information retrieval~\cite{hoffer2015deep}.

Recently, Triplet loss has been proposed for speech tasks, for instance: the speaker verification task ~\cite{zhang2018text}, for speaker turn~\cite{bredin2017tristounet}, speech emotion classification~\cite{huang2018speech}, among other tasks. However, its applicability to forensic speaker identification has not been explored, particularly for the Spanish language. \emph{Forensic Speaker Identification} (FSI) focuses on gathering and quantifying the evidence that will be presented in a court. FSI addresses the question if a specific recording registers or not, speech produced by a specific person~\cite{rose2002forensic,morrison2019forensic}. The more basic scenario in FSI consists of two sample speech recordings, a reference sample and a questioned sample. For the reference sample, we always know the identity of the speaker. This certainty is guaranty by the chain of custody, this is because we know the conditions in which the recording was taken, including the identity of the speaker. On the other hand, for the questioned recording we are not sure about the identity of the person whose voice is in the recording. In a case that involves FSI, the identity of the speaker in the questioned recording is contested regarding the identity in the reference recording; one of the involved parts affirms that the voice in the reference and the questioned recordings are the same (\emph{same-speaker hypothesis}); while the other part affirms the contrary (\emph{different-speaker hypothesis}). 

The goal in a FSI case, is not only about matching the two recordings by their \emph{similarity}, same-speaker hypothesis, FSI requires a stronger legal standard which makes also necessary to quantify the chances of the questioned sample to be associated to other speakers of the population, this address the different-speaker hypothesis. This measurement is known as \emph{typicality}. With these two measurements, similarity and typicality it is common to calculate the Likelihood Radio (LR)~\cite{good1991weight}. LR offers a quantifiable measurement that updates the odds of one of the hypothesis, this information should be taken into consideration in the context of the presence of other evidence regarding the case.  

In this work, we propose to extract speech embeddings from speech spectrogram samples for reference and questioned speech recordings in order to quantify the LR. We start by presenting related work in section~\ref{sec:rw}. We continue by presenting the details of our neural model and the implementation of the triplet loss in section~\ref{sec:nm}. We present two sets of results: first, in section~\ref{exp1}, we measure the quality of the embeddings by proposing inner and outer speaker distance metrics, together with the use of the well established silhouette clustering metric. Second, in section~\ref{exp2}, we propose two ways to calculate the LR in terms of the speech embedding distances. Once we show that speech embeddings are an option to be used in FSI we discuss some ethical aspects to be considered in section~\ref{sec:ethics}. Finally, we summarise our main findings in section~\ref{sec:con}.

\section{Related work}
\label{sec:rw}

Forensic speech science has advanced in last three decades in which it has established different methodologies and techniques to face the speech identification problem~\cite{morrison2019forensic}. In the case of the methodology, there has been a paradigm shift towards empirically grounded methods~\cite{saks2005coming,morrison2009forensic}. This shift has been motivated by the requirements of admissibility of science evidence that had become a standard in some courts around the world~\cite{giannelli1980admissibility,champod2011scientific}. The main result of this shift has been the adoption of the Likelihood-Ratio (LR) as a mean to introduce the evidence in court. LR is formulated in the following manner:
\begin{equation}
    LR=\frac{p(E|H_s)}{p(E|H_d)}
\end{equation}
where $E$ represents the evidence, in FSI this is the quantification of speech properties in the questioned recorded sample. $H_s$ correspond to the same-speaker hypothesis and $H_d$ to the different-speaker hypothesis. The numerator can be considered a similarity score while the denominator a typicality one. LR is not to be considered independent of the other facts of the case, on the contrary its meaning only depends on the strength or lack of the rest of the evidence and their compatibility with either of the hypotheses. 

On the other hand, from the point of the techniques there has been several proposals which allow the quantification of $p(E|{H_s})$ and $p(E|{H_d})$. One approach that was extensively explored was the statistical analysis supposing a Gaussian distribution of speech features~\cite{rose2002forensic}. A common approach is to measure a specific phonetic and phonological speech properties (e.g., formants) in a specific context (e.g., a word \cite{rose1998forensic,rose1999long}). Motivated by the application of multi-variable statistics in the forensic field~\cite{aitken2004evaluation} new approaches were suggested together with kernel approaches to improve the statistical analysis of the speech evidence~\cite{rose2004linguistic,morrison2011comparison}.

Another proposed method is the use Gausian-Mixture-Model/Universal Background Model (GMM-UBM)~\cite{bimbot2004tutorial}. This methodology reflects more a generative machine learning approach which depends on data intensive algorithms, for this reason in these methodologies there is no necessity of measuring a specific voice property. The GMM-UBM model is parametric deppends on a dataset of recordings which is used to define the model parameters, in this case the parameters of the  Gaussian mixtures. A GMM speaker-specific model is created to quantify the similarity term, and a UBM general model, based on the population of possible speakers, is generated to quantify the typicality term. With the advent of different machine learning techniques some other machine learning-based approaches had been proposed, such as using Support Vector Machines, boosting algorithms and Random Forest~\cite{univaso2015data,morrison2020statistical}. A fundamental piece to adopt ML techniques is its appropriateness to calculate the LR~\cite{1202302}. Recently, there has been proposals that exploits the discriminative power of Neuronal Networks and their capability of producing representations, some examples of these approaches are: DNN senone i-vectors~\cite{garcia2014supervised}, bottleneck features~\cite{yaman2012bottleneck} and x-vectors~\cite{snyder2018x}.

On the other hand, with the progress of self-supervised methods for training deep learning networks there has been advances in proposals for learning representations for embedding spaces. Contrastive methods have been proposed to compare a questioned sample to a set of samples of known speakers, this setting correspond to a setting of \emph{speaker verification} were there is access to samples of the speaker to identify, it is a matter of verifying if the two sets of samples match~\cite{heigold2016end,snyder2018x}.  Of particular interest are the advances with the Triplet Loss, since it maps raw speech representation into an euclidean space~\cite{schroff2015facenet}. The use of these embeddings for the speaker identification task is trivial since the distance among embeddings can be used to determine which speaker is close to a questioned recording. Here it is important to notice, that speaker identification does not quantify typicality, but just similarity. This approach had been used in different scenarios: \cite{li2017deep} uses a Residual CNN and GRU that transforms a spectogram into a $512$ dimension vector, it proposes to use cosine similarity to guide the triplet loss and it evaluates accuracy and error to recognise the speaker. \cite{zhang2017end} presented a CNN Inception Resnet to generate a $128$ dimension embedding, it focused on $L2$ norm, it proposed validation and false accept rate to evaluate its system. \cite{mo2020self} modifies the triplet loss to make it more efficient, it also uses accuracy (top 1 and top 5) to evaluate its system. In all these three cases, the resulting performances were superior to previous approaches. 

Of particular interest for our experimentation is the difference between female and male voices, since as it will be presented further down that our dataset has an imbalance among this type of speakers. According to the medical notion, the voice originates in the throat of the speaker specifically in the larynx. There is an understanding that the size of the larynx correlates with sexual characteristics which at the same time determine the sex (or pitch) of the voice as an acoustic event. In average, the male larynx is larger than the female larynx and is naturally inclined to produce a lower pitched speech. However this is not a rule since for men and women speech can overlap. In this regard, the sexual characteristics of the vocal tract, biologically determined, determine sex in speech, however the identification phase is of one of gender, since it is constituted by a subjective interpretation~\cite{azul2013voices,mcdermott2011distinguishing,ertam2019effective}. With this in mind our experimentation will be on perception which means on gender.

\section{Triplet loss and Neural Model}
\label{sec:nm}

\begin{figure}[t]
    \centering
    \fontsize{5}{5}\selectfont
    \includegraphics[width=0.8\textwidth]{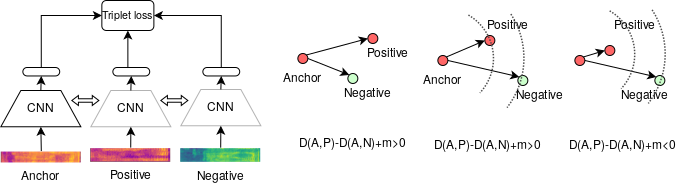}
    \caption{Triplet loss applied to three speech segments which are transformed into three vectors. Possible cases relation among A, P and N inputs, the third case is a zero loss case.}
    \label{fig:tripletloss}
\end{figure}

Triplet loss compares an anchor input with two other inputs, a positive input which shares a property with the anchor, in our case it is the identity of the speaker, and with a negative input which does not share such property. The comparison is guided by the following formulation:
\begin{equation}
    L(A,P,N) = max( D(A,P) - D(A,N) + m, 0)
\end{equation}
where $A$, $P$ and $N$ are vectors representing the anchor, the positive and negative inputs respectively. $D$ is distance metric and $m$ is a margin. In a ideal setting it is expected that the distance between $A$ and $P$ to be less than distance between $A$ and $N$ at least by a margin $m$, if that is the case the loss is zero when the network calculating the vectors is doing a good job. However, if this is not the case the loss will be positive and by using back propagation the weights of the model that produces the vectors from raw information will be adjusted. Figure~\ref{fig:tripletloss} shows the relation between the CNN model and the triplet loss, it is important to notice that the CNN blocks are the same neuronal network that transforms all inputs since the weights are shared. The Figure also illustrates the three cases in relation to distances among the embeddings and the margin. $L2$ norm is used as a distance metric.

\begin{figure}[t]
    \centering
    \fontsize{5}{5}\selectfont
    \includegraphics[width=0.8\textwidth]{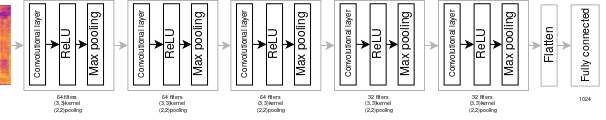}
    \caption{Convolutional network}
    \label{fig:cnn}
\end{figure}

A common arrangement in Triplet loss is to use the same neuronal model to produce the vector inputs from raw inputs. In this work we propose the use a CNN that will receive segments of the spectogram of speech recordings. Figure~\ref{fig:cnn} shows this arrangement. We propose a simple CNN composed by $5$ layers when possible otherwise $4$. Figure~\ref{fig:cnn} shows a diagram with the specific details of our model. 

\subsection{Preprocessing of speech signal}

The input of our CNN is a segment of a spectogram that represents a time slice ($t$ in milliseconds) and frequencies information up to $8.5 kHz$ (enough to characterise the human voice), using always $256$ bins for the frequencies. All our recordings are down sampled to $16$K and the spectogram is normalized and pre-amph. These specific choices are followed from typical pre-processing of speech signals. In order to obtain the spectogram we use a \emph{hann} window, with variable window ($w$ in milliseconds) and hop ($h$ in milliseconds) size. The parameters $t$, $w$ and $h$ allows us to generate an image patch (spectogram segment) with a variable width but a constant height $W\times 256$. This patch is the CNN's raw input which will be transformed it into a $1024$ dimension embedding for all our experiments.

\section{Dataset}
\label{sec:dataset}

In this work we use the Spanish Voxforge dataset, which is entirely based on the recordings from the Voxforge Project\footnote{\url{http://www.voxforge.org}}. The Voxforge Project is a non-profit initiative that aims to collect transcribed speech for use with Free and Open Source Speech Recognition Engines. We chose the Voxforge because the speakers read always the same prompt, a paragraph of \emph{El Quijote}, this eliminates overfitting by the content of what it is said. We expect that our models focus on properties of how things are said. The speakers donated a sample of their voice by registering to the project website, filling a form with relevant information about the speaker and reading some prompts directly through the speaker's computer microphone. Thanks to this mechanism, we can know the following about every speaker, which is relevant for forensic purposes and our experimentation:

\begin{itemize}
    \item Username: It could be left in blank or it could be an alias
    \item Gender: Male / Female
    \item Age: $13-17=$ Youth / $18-64=$ Adult / $65$-or greater$=$ Senior)
    \item Native Speaker?: Yes / No
    \item Dialect: Country or Region
\end{itemize}

Since the beginning of the project in $2006$, several languages have been added by the community, making necessary to clarify that our corpus only contains Spanish recordings until the year $2016$. The original recordings were manually segmented into utterances. Table~\ref{tbl:nationality} shows the total number of male and female speakers and how they are classified into the corpus according to their nationality.

\begin{table}[ht]
    \centering
    \begin{tabular}{|l|c|c|}\hline
         Country       &  Males & Females\\\hline
         Argentina     &  $143$ &  $31$  \\
         Chile         &   $69$ &   $3$  \\
         Latin America &  $148$ &  $10$  \\
         Mexico        &   $68$ &   $8$  \\
         Spain         & $1240$ & $411$  \\
         Unknown       &  $ 45$ &   $4$  \\\hline
         Total         & $1713$ & $467$  \\\hline
    \end{tabular}
    \caption{Number of Speakers in the Voxforge Corpus and their Nationalities.}
    \label{tbl:nationality}
\end{table}

The Spanish Voxforge dataset is composed by $21,692$ recordings which in average last $8.25$ seconds. They come in a $16kHz$, $16$ bit, mono format. The total duration of the whole recordings is approximately $50$ hours. For the experiments, we split the speakers in $80\%$ training, $10\%$ validation and $10\%$ testing.

\section{Experiments and results}

We performed two levels of experimentation. First we explored the speech embeddings by measuring quality of the speech embeddings and the cluster they generate on samples from the same spaker. Second we evaluated the capability of the speech embeddings to be used to calculate an LR, we propose two approaches: one is to use directly distance and questioned as a proxy for the LR, second is to use a ratio between the previous distance divided by the shortest distance to other speaker in the population.

\subsection{Quality of the embeddings}
\label{exp1}

To quantify the quality of the embeddings, we calculate three metrics: inner average distance for the same speaker samples ($IAD$), outer average distance between speaker and centroids of other speakers ($OAD$). We expect a IAD small which will signal that samples from the same speaker will land in the same region. For OAD, we expect a large number that will signal that the samples from a different speaker will be far away from other speakers. With these two metrics we proposed to calculate a distance ratio ($DR$) that will tell us the relation between a speaker and the rest of speakers, we will expect a small amount for $DR$ signaling good speech embeddings. We also calculated the  mean silhouette coefficient ($MSC$) which ranges from $-1$ (worst result) to $1$ (best result). In this case, a number closer to $1$ will indicate that there is less confusion among the clusters of embeddings from the same speaker samples. 

The first question we address is how long we have to train the network using triplet loss. For these experiments we set the parameters to $t=2000ms$, $w=100$ and $h=50$, we also set a margin of $2$ and we train the model during $1$, $2$ and $3$ days  (exploratory experiments with shorter time showed that the minimum training time was a day). Table~\ref{tbl:exp1_time} reports the results. As can be seen $1$ day of training is enough to reach a good results:

\begin{table}[ht]
    \centering
    \begin{tabular}{|c||c|c|c|c|c|}\hline
    Training days & IAD & IOD & DR & MSC \\\hline
    1 & $4.77$ & $27.61$ & $\textbf{0.1730}$ & $\textbf{0.2475}$ \\
    2 & $\textbf{4.61}$ & $26.62 $ & $0.1732 $ & $0.2326$ \\
    3 & $4.85$ & $\textbf{27.88}$ & $0.1742$ & $0.2358$ \\\hline
    \end{tabular}
    \caption{Validation results for different training duration, one day gives the best results  amd more time does not affect the behaviour of the network. }
    \label{tbl:exp1_time}
\end{table}

The second question we address was how large has to be the CNN's input speech segment. For this we have to explore different parameters of the spectogram: $h$, $t$ while we fix $w$ to $100$ ms which is a common window size for speech signal processing. Varying $h$ allows us to control the amount of information that passes through in a segment, we explored $25$ms, $33$ms and $50$ms. On the other hand $t$ allows to control the amount of signal that the CNN will 'see', we explore $1$s, $1.5$s and $2$s values; we did not try a larger time since some all of recordings were at least $2$s, but not necessary longer. We set the margin to $2$ since in previous experiments had allows to identify a good compromise between $DR$ and $MSC$. Table~\ref{tbl:exp1_all} summarise the main findings for different combinations of these parameters, each model was trained during a day. As can be seen the more information in the patch the smaller the DR is (which is a good result). This points that a more informative patch is obtained by a larger segment with a small hope size. 

\begin{figure}[ht]
    \centering
    \includegraphics[width=0.45\textwidth]{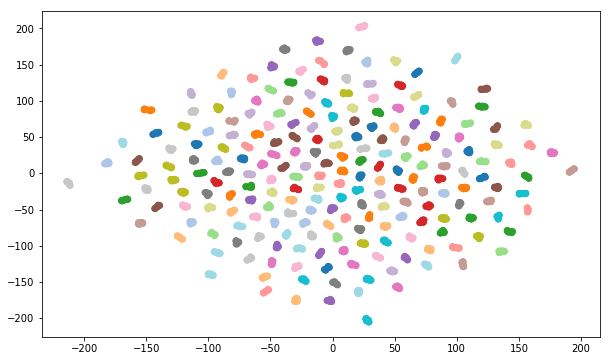}
    \includegraphics[width=0.45\textwidth]{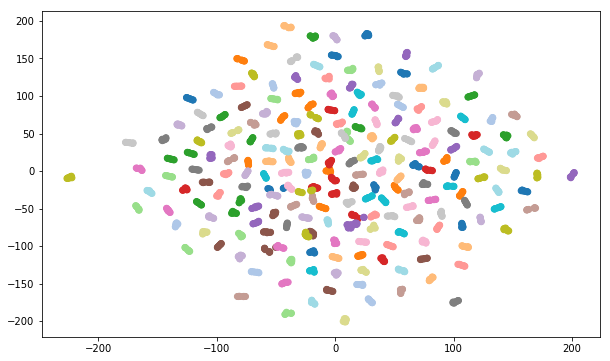}
    \caption{Projections of a recording for speakaer in validation set. First projection corresponds to our best model ($t=2000$, $h=25$), second to our worst model ($t=1000$, $h=25$) (same color same speaker, 218 speakers)}
    \label{fig:projection}
\end{figure}

Figure~\ref{fig:projection} shows the projection of embeddings. As it can be seen same speaker embeddings clusters, in this case these cluster represent one speaker's recording from which we extracted $20$ samples that became $20$ embeddings. The figure illustrates our best ($t=2000$ and $h=25$) and worst models ($t=1000$ and $h=50$). As it can be seen in both projections there is ordering of the speakers (clusters of speakers) but also shows some speakers which are close among themselves. It can been seen that our best model produces more organized positions while for the worst model there is some confusion among the clusters in the middle.  

\begin{table}[ht]
    \centering
    \begin{tabular}{|c|c||c|c|c|c|c|}\hline
    $t$ (ms) & $h$ (ms)  & Patch size & IAD & IOD & DR & MSC \\\hline
    $1000$ & 25 & $40\times 256$  & $3.21$ & $22.17$ & $\textbf{0.1445}$ & $\textbf{0.1924}$ \\
    $1000$ & 33  & $30\times 256$* & $3.60$ & $19.36$ & $0.1860$ & $0.1754$ \\
    $1000$ & 50  & $20\times 256$* & $4.68$ & $23.21$ & $0.2018$ & $0.1506$ \\\hline
    $1500$ & 25  & $60\times 256$  & $3.57 $ & $25.74$ & $\textbf{0.1386} $ & $ \textbf{0.2209}$ \\
    $1500$ & 33  & $45\times 256$  & $3.42$ & $21.31$ & $0.1605$ & $0.2276$ \\
    $1500$ & 50 & $30\times 256$* & $3.98$ & $23.46$ & $0.1695$ & $0.1721$ \\\hline
    $2000$ & 25& $80\times 256$  & $4.69$ & $30.24$ & $0.1541$ & $\textbf{0.2766}$ \\
    $2000$ & 33  & $60\times 256$  & $4.44$ & $29.70$ & $\textbf{0.1495}$ & $0.2487$ \\
    $2000$ & 50  & $40\times 256$  & $4.77$ & $27.60$ & $0.2019$ & $0.2475$ \\\hline
    \end{tabular}
    \caption{Validation results for different parameters of speech segments. }
    \label{tbl:exp1_all}
\end{table}

\begin{table}[ht]
    \centering
    \begin{tabular}{|c|c|}\hline
    Metric & Value\\\hline
    IAD & $4.46$\\
    IOD & $28.68$\\
    DR  & $0.1555$\\
    MSC & $0.2618$\\\hline
    \end{tabular}
    \caption{Evaluation results with parameters \textit{t=2000}, \textit{h=25}, \textit{w=100} and \textit{m=2}. }
    \label{tbl:exp1_final}
\end{table}

As presented in the section \ref{sec:dataset} there is an imbalance between female and male speakers, and between the nationalities of the speakers. To quantify the effect of these imbalances we performed more evaluations. In the case of gender we created three models: with only female recordings ($F$, $373$ training and $47$ validation speakers), with only male recordings ($M*$,$1370/171$) and  training with a comparable amount of male recordings with female recordings ($M$, $373/47$). For this experiments we set the parameters to $t=2000ms$, $w=100ms$ and $h=50ms$. Table~\ref{tbl:gender} shows the main results on the corresponding validation dataset. As it can be seen there is an effect of gender miss-match, however this is not severe for the case of training with female speakers and using the model with male speakers. On the other direction, we can notice a severe drop in performance. As expected, the best setting is to train with the biggest amount of recordings of a gender and evaluating on that gender. 

\begin{table}[ht]
    \centering
    \begin{tabular}{|c|c|||l|l|l|l|l|}\hline
    Model & Evaluating & IAD & IOD & DR & MSC \\\hline
    $F$  & $F$ & $3.74$ & $22.23$  & $0.1686$ & $0.2500$ \\
    $F$  & $M$ & $4.68$ & $19.84$  & $0.2357$ & $0.2571$ \\\hline
    $M$  & $M$ & $6.29$ & $33.17$  & $0.1896$ & $0.3517$ \\
    $M$  & $F$ & $5.97$ & $33.12$  & $0.1822$ & $0.2381$ \\\hline
    $M$* & $M$ & $3.71$ & $21.03 $ & $0.1765 $ & $0.3236 $ \\
    $M$* & $F$ & $3.32$ & $22.87 $ & $0.1454 $ & $0.2073 $ \\\hline
    \end{tabular}
    \caption{Evaluations measure the effect of gender imbalance, $F$ (female) and $M$  (male) are comparable since they have the same number of training speakers; $M$* is not directly comparable since it relies in a larger amount of speakers.}
    \label{tbl:gender}
\end{table}

Table~\ref{tbl:nationallity} shows our findings for the nationality. Given the amount of speakers, we decided to compare two types of speakers from two regions: Latinamerica ($L$) and Spain ($S$). We set two models, $L$ ($373$/$48$) and  $S$ ($373$/$48$). As we can see there is not a notable difference given the nationality, considering the region Latinamerica packs more nationalities in the dataset we can not see an effect of the regional accents on the capabilities of triplet loss in producing good embeddings. Similarly to gender, the miss-match between training and evaluation datasets speakers produces drops in the performance, but not as severe as one might expect. 

\begin{table}[ht]
    \centering
    \begin{tabular}{|c|c|||l|l|l|l|l|}\hline
    Model & Evaluating & IAD & IOD & DR & MSC \\\hline
    $L$  & $L$ & $5.05$ & $21.73$  & $0.2325$ & $0.3410$ \\
    $L$  & $S$ & $4.68$ & $25.17$  & $0.1903$ & $0.3401$ \\\hline
    $S$  & $S$ & $4.41$ & $26.42$  & $0.1670$ & $0.3517$ \\
    $S$  & $L$ & $4.73$ & $20.10$  & $0.2355$ & $0.2911$ \\\hline
    \end{tabular}
    \caption{Evaluations measure the effect of nationality imbalance, $L$ (Latinamerica) and $S$ (Spain)}
    \label{tbl:nationallity}
\end{table}

\subsection{Speech embeddings for FSI}
\label{exp2}

In these experiments we aim to establish a way to calculate the LR based on the distances among embeddings. The more straight forward proposal is to use a normalized distance between the centroids of the reference and questioned sample embeddings, we call this approach distance based ($D$). The second proposal is to use a radio between the same distance $D$ and the distance to the closest speaker in the population in order to account for tipicallity, we call this approach distance radio ($DR$). The first proposal can be formalized as:
\begin{equation}
    LR_{D} = \frac{D(q,r)}{N};
\end{equation}
while the second proposal as:
\begin{equation}
    LR_{DR} = \frac{min(D(q,p))}{D(r,q)};\forall p\in Population
\end{equation}
Where $q$ is the centroid of the questioned embedding samples, $r$ the centroid of the reference and $p$ is a centroid of a speaker from the population and $N$ is a normalizing factor.

\begin{figure}[ht]
    \centering
    \includegraphics[width=0.5\textwidth]{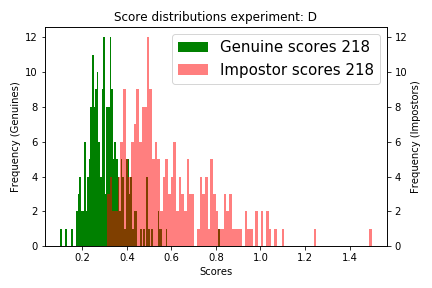}
    \includegraphics[width=0.5\textwidth]{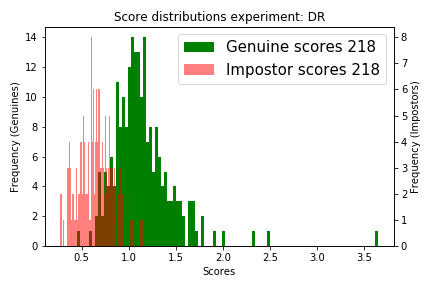}
    \caption{Histogram of scores for positive (Genuine scores) and negative (Impostor scores) for both LR cases}
    \label{fig:exp2_freqdet}
\end{figure}

\begin{figure}[ht]
    \centering
    \includegraphics[width=0.50\textwidth]{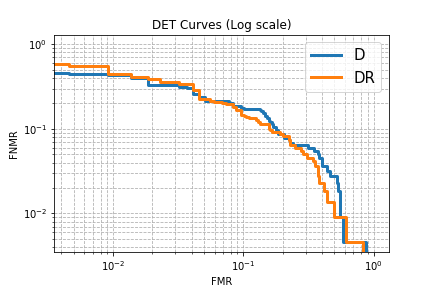}
    \includegraphics[width=0.50\textwidth]{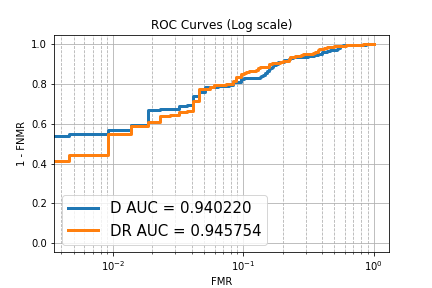}
    \caption{DET and ROC Log curves, for both proposals:  distance (D) and distance ratio (DR).}
    \label{fig:exp2_comparison}
\end{figure}

For the experimentation we used two Forensic Speech Identification scenarios: genuine (were the same-speaker hypothesis is true) and impostor (where the different-speaker hypothesis is true). Per speaker we randomly selected three recordings from where we sampled $15$ segments per recording as a reference. In the case of the \emph{genuine} scenario we selected a fourth recording as the questioned source of samples while for the impostor scenario we randomly selected an extra recording from a different speaker. Additionally, as our population we randomly selected $100$ different speakers from which we extracted the same amount of samples that our reference (i.e., $45$). It is important to remark that the recordings for these settings came from the validation split of the data.
With this considerations, we have have $218$ genuine cases and $218$ impostor cases. Figure~\ref{fig:exp2_freqdet} shows for both approaches the LR scores: distance based ($D$) and distance radio ($DR$). For the case of the distance as a proxy for LR ($D$), as expected genuine cases are concentrated with lower values with a mean in $0.31$ while impostor are located with a mean of $0.59$. The Equal Error Rate (EER) for the metric is of $0.1467$ with a sensitivity of $1.80$.  For the case of distance radio ($DR$) we see that it corresponds better with the common interpretation of LR were the score indicates an update in the believe on the same speaker hypothesis. In this case, lower scores are associated to impostor cases with a mean of $0.64$  and larger score correspond to genuine cases with a mean of $1.15$. This approach has an EER $0.13$ and sensitivity index of $1.79$.
\newpage
Figure~\ref{fig:exp2_comparison} shows the DET and ROC curves in a log scale for both scores. DET curves are common in the forensic field, in particular it compares the False Match Rate (FMR) with False Non-Match Rates (FNMR) of the system. For both DET curves we see that as the number of false positives grows it is harder to missmatch a case. A better system  produces a curve located to the left bottom corner. Our experiments show that bot approaches $D$ and $DR$ share a lot of predictive power. It is important to take into account that DET curves are related to ROC curves, which are more common in the ML field. Within the ROC curves is common to measure Area Under the Curve (AUC) to compare two systems, in this case we can observe that the $DR$ approach has AUC of $0.9475$ while $D$ a AUC score of $0.9402$, this points to a slightly more robust discriminative power for the $DR$ metric.
  
%---------------------------------------------------------------%

\section{Ethical discussion}
\label{sec:ethics}

In recent years, the topic of sharing personal information has become extremely relevant due to the intensive interaction between modern societies and technology, in particular technology that are able to take autonomous decisions. Nowadays, simple things like clicking a ``like'' button or typing some words to search in a search engine could trigger undesired advertisements or the risk to be scammed in creative ways~\cite{irani2011modeling}. As computer scientists, these type of observations raise the question of how datasets are share through the internet in order to contribute to the advances of science but at the same time, not to harm anyone in anyway, as it can be understood in the \emph{hippocratic oath for artificial intelligence practitioners}~\cite{etzioni2018hippocratic}. 

In this work, the use of the Voxforge Spanish dataset raises some privacy concerns, since the donors accepted that their voices are part of a database destined to create language technologies however while maintaining anonymity. On the other hand, due to the type of information provided by each donor (mainly a personalised username), it could be easy to identify some of them and performing actions against them like identity fraud in certain speech systems~\cite{safavi2016fraud}. Fortunately, some of this concerns were contemplated by the developers of the Voxforge datasets which allowed and option to opt-out, however it could hardly be enforced for all distributed copies~\cite{zimmer2010but}. Following this concern our experiments do not rely on the \emph{username} field and it only releases the models, not the recordings. We also will honor any opt-out petition to the Voxforge dataset.

Other aspect of concern regarding this work is if the proposed system is ready to be deployed and used in a real court case. From our perspective, even though the experiments show an interesting and promising performance, it is clear that our models are not ready for deployment in court. In particular, the selection of the population has to be properly. In a real case, it must be transparent who will be included in the population to calculate typicality. Additionally, the results of our models have to be used in the context of other evidence, as previously mentioned the LR has the goal of updating the odds of a legal hypothesis, but it is not a definitive test of identity. So in this context, any attempt to use our models as a recognition system will be ill-advised.

%---------------------------------------------------------------%

\section{Conclusions}
\label{sec:con}

In this work we explored the use of the triplet loss function and a convolution neural network (CNN) for the forensic identification of speakers. We have shown that this setting offers an alternative to well established approaches. Within this setting, the goal is to train a convolutional network (CNN) to produce an embedding of a speech sample. Although previous research have shown that such embeddings can cluster speakers, in our experimentation we have shown that the distance among clusters can be used to approximate a Likehood Radio (LR) which is a common measurement to convey the plausibility of a legal forensic hypothesis. 

Our experimentation points out that the CNN is able to produce a good embedding representation for long speech samples and finer resolution of the spectogram. We hypothesize this is because it contains more information of a speaker voice. In particular we have focused on Spanish from Latin America and Spain, we have shown that our approach is not affected by variant of Spanish. However, gender has a strong effect on the performance. So far our results points out that models have to be gender dependent, but more research has to be done on this point (see~\cite{ertam2019effective}).
 
Finally, our results also showed that triplet loss conveys a good  discriminative power when the distances are used to approximate LR.  We proposed two approaches for the calculation of LR, the first one is solely based on the distances among a questioned and reference samples ($D$). This approach can be considered independent of the population. The second approach uses the radio between the first approximation $D$ and the distance of the questioned sample to the closest speaker from the population ($DR$). We conclude that this second option provides a better alternative, since it conveys information in two aspects: first numerically it corresponds to other LR scores and second it has a better predictive power (AUC=$0.9458$), this makes it suitable for the forensic identification of speakers.

\section{Acknowledgements}
The authors thank CONACYT for the computer resources provided through the
INAOE Supercomputing Laboratory's Deep Learning Platform for Language
Technologies with the project \emph{Experimentos con voz, traducci\'on y clasificaci\'on de textos} (id. PAPTL 01-008). We also acknowledge Fernanda Hernandez and Sandra V\'azquez which were involved in early stages of the development of the code and experimentation.

% Authors must disclose all relationships or interests that 
% could have direct or potential influence or impart bias on 
% the work: 
%
\section*{Conflict of interest}

The authors declare that they have no conflict of interest.

\bibliographystyle{unsrt}  
%\bibliography{references}  %%% Remove comment to use the external .bib file (using bibtex).
%%% and comment out the ``thebibliography'' section.

%%% Comment out this section when you \bibliography{references} is enabled.
% BibTeX users please use one of
%\bibliographystyle{spbasic}      % basic style, author-year citations
%\bibliographystyle{spmpsci}      % mathematics and physical sciences
%\bibliographystyle{spphys}       % APS-like style for physics
\bibliography{references}   % name your BibTeX data base

\begin{thebibliography}{10}

\bibitem{rose1997identifying}
Philip Rose.
\newblock Identifying criminals by their voice-the emerging applied discipline
  of forensic phonetics.
\newblock {\em Australian Language Matters}, 5(2):6--7, 1997.

\bibitem{schroff2015facenet}
Florian Schroff, Dmitry Kalenichenko, and James Philbin.
\newblock Facenet: A unified embedding for face recognition and clustering.
\newblock In {\em Proceedings of the IEEE conference on computer vision and
  pattern recognition}, pages 815--823, 2015.

\bibitem{wang2014learning}
Jiang Wang, Yang Song, Thomas Leung, Chuck Rosenberg, Jingbin Wang, James
  Philbin, Bo~Chen, and Ying Wu.
\newblock Learning fine-grained image similarity with deep ranking.
\newblock In {\em Proceedings of the IEEE Conference on Computer Vision and
  Pattern Recognition}, pages 1386--1393, 2014.

\bibitem{cheng2016person}
De~Cheng, Yihong Gong, Sanping Zhou, Jinjun Wang, and Nanning Zheng.
\newblock Person re-identification by multi-channel parts-based cnn with
  improved triplet loss function.
\newblock In {\em Proceedings of the iEEE conference on computer vision and
  pattern recognition}, pages 1335--1344, 2016.

\bibitem{hoffer2015deep}
Elad Hoffer and Nir Ailon.
\newblock Deep metric learning using triplet network.
\newblock In {\em International Workshop on Similarity-Based Pattern
  Recognition}, pages 84--92. Springer, 2015.

\bibitem{zhang2018text}
Chunlei Zhang, Kazuhito Koishida, and John~HL Hansen.
\newblock Text-independent speaker verification based on triplet convolutional
  neural network embeddings.
\newblock {\em IEEE/ACM Transactions on Audio, Speech, and Language
  Processing}, 26(9):1633--1644, 2018.

\bibitem{bredin2017tristounet}
Herv{\'e} Bredin.
\newblock Tristounet: triplet loss for speaker turn embedding.
\newblock In {\em 2017 IEEE international conference on acoustics, speech and
  signal processing (ICASSP)}, pages 5430--5434. IEEE, 2017.

\bibitem{huang2018speech}
Jian Huang, Ya~Li, Jianhua Tao, Zhen Lian, et~al.
\newblock Speech emotion recognition from variable-length inputs with triplet
  loss function.
\newblock In {\em Interspeech}, pages 3673--3677, 2018.

\bibitem{rose2002forensic}
Phil Rose.
\newblock {\em Forensic speaker identification}.
\newblock cRc Press, 2002.

\bibitem{morrison2019forensic}
Geoffrey~Stewart Morrison, Cuiling Zhang, and Ewald Enzinger.
\newblock Forensic speech science.
\newblock {\em The Bloomsbury companion to phonetics}, pages 183--197, 2019.

\bibitem{good1991weight}
Irving~J Good.
\newblock Weight of evidence and the bayesian likelihood ratio.
\newblock {\em The use of statistics in forensic science}, pages 85--106, 1991.

\bibitem{saks2005coming}
Michael~J Saks and Jonathan~J Koehler.
\newblock The coming paradigm shift in forensic identification science.
\newblock {\em Science}, 309(5736):892--895, 2005.

\bibitem{morrison2009forensic}
Geoffrey~Stewart Morrison.
\newblock Forensic voice comparison and the paradigm shift.
\newblock {\em Science \& Justice}, 49(4):298--308, 2009.

\bibitem{giannelli1980admissibility}
Paul~C Giannelli.
\newblock The admissibility of novel scientific evidence: Frye v. united
  states, a half-century later.
\newblock {\em Colum. L. Rev.}, 80:1197, 1980.

\bibitem{champod2011scientific}
Christophe Champod and Jo{\"e}lle Vuille.
\newblock Scientific evidence in europe--admissibility, evaluation and equality
  of arms.
\newblock {\em International Commentary on Evidence}, 9(1), 2011.

\bibitem{rose1998forensic}
Phil Rose.
\newblock A forensic phonetic investigation into non-contemporaneous variation
  in the f-pattern of similar-sounding speakers.
\newblock In {\em Fifth International Conference on Spoken Language
  Processing}, 1998.

\bibitem{rose1999long}
Phil Rose.
\newblock Long-and short-term within-speaker differences in the formants of
  australian" hello".
\newblock {\em Journal of the international Phonetic Association}, pages 1--31,
  1999.

\bibitem{aitken2004evaluation}
Colin~GG Aitken and David Lucy.
\newblock Evaluation of trace evidence in the form of multivariate data.
\newblock {\em Journal of the Royal Statistical Society: Series C (Applied
  Statistics)}, 53(1):109--122, 2004.

\bibitem{rose2004linguistic}
Philip Rose, D~Lucy, Takashi Osanai, et~al.
\newblock Linguistic-acoustic forensic speaker identification with likelihood
  ratios from a multivariate hierarchical random effects model-a non-idiot's
  bayes' approach.
\newblock {\em Proceedings of the 10th Australian International Conference on
  Speech Science and Technology}, 2004.

\bibitem{morrison2011comparison}
Geoffrey~Stewart Morrison.
\newblock A comparison of procedures for the calculation of forensic likelihood
  ratios from acoustic--phonetic data: Multivariate kernel density (mvkd)
  versus gaussian mixture model--universal background model (gmm--ubm).
\newblock {\em Speech Communication}, 53(2):242--256, 2011.

\bibitem{bimbot2004tutorial}
Fr{\'e}d{\'e}ric Bimbot, Jean-Fran{\c{c}}ois Bonastre, Corinne Fredouille,
  Guillaume Gravier, Ivan Magrin-Chagnolleau, Sylvain Meignier, Teva Merlin,
  Javier Ortega-Garc{\'\i}a, Dijana Petrovska-Delacr{\'e}taz, and Douglas~A
  Reynolds.
\newblock A tutorial on text-independent speaker verification.
\newblock {\em EURASIP Journal on Advances in Signal Processing},
  2004(4):101962, 2004.

\bibitem{univaso2015data}
Pedro Univaso, Juan~Maria Ale, and Jorge~A Gurlekian.
\newblock Data mining applied to forensic speaker identification.
\newblock {\em IEEE Latin America Transactions}, 13(4):1098--1111, 2015.

\bibitem{morrison2020statistical}
Geoffrey~Stewart Morrison, Ewald Enzinger, Daniel Ramos, Joaqu{\'\i}n
  Gonz{\'a}lez-Rodr{\'\i}guez, and Alicia Lozano-D{\'\i}ez.
\newblock {\em Statistical models in forensic voice comparison}.
\newblock CRC Press LLC Boca Raton, Florida, 2020.

\bibitem{1202302}
J.~{Gonzalez-Rodriguez}, J.~{Fierrez-Aguilar}, and J.~{Ortega-Garcia}.
\newblock Forensic identification reporting using automatic speaker recognition
  systems.
\newblock In {\em 2003 IEEE International Conference on Acoustics, Speech, and
  Signal Processing, 2003. Proceedings. (ICASSP '03).}, volume~2, pages II--93,
  2003.

\bibitem{garcia2014supervised}
Daniel Garcia-Romero and Alan McCree.
\newblock Supervised domain adaptation for i-vector based speaker recognition.
\newblock In {\em 2014 IEEE International Conference on Acoustics, Speech and
  Signal Processing (ICASSP)}, pages 4047--4051. IEEE, 2014.

\bibitem{yaman2012bottleneck}
Sibel Yaman, Jason Pelecanos, and Ruhi Sarikaya.
\newblock Bottleneck features for speaker recognition.
\newblock In {\em Odyssey 2012-The Speaker and Language Recognition Workshop},
  2012.

\bibitem{snyder2018x}
David Snyder, Daniel Garcia-Romero, Gregory Sell, Daniel Povey, and Sanjeev
  Khudanpur.
\newblock X-vectors: Robust dnn embeddings for speaker recognition.
\newblock In {\em 2018 IEEE International Conference on Acoustics, Speech and
  Signal Processing (ICASSP)}, pages 5329--5333. IEEE, 2018.

\bibitem{heigold2016end}
Georg Heigold, Ignacio Moreno, Samy Bengio, and Noam Shazeer.
\newblock End-to-end text-dependent speaker verification.
\newblock In {\em 2016 IEEE International Conference on Acoustics, Speech and
  Signal Processing (ICASSP)}, pages 5115--5119. IEEE, 2016.

\bibitem{li2017deep}
Chao Li, Xiaokong Ma, Bing Jiang, Xiangang Li, Xuewei Zhang, Xiao Liu, Ying
  Cao, Ajay Kannan, and Zhenyao Zhu.
\newblock Deep speaker: an end-to-end neural speaker embedding system.
\newblock {\em arXiv preprint arXiv:1705.02304}, 2017.

\bibitem{zhang2017end}
Chunlei Zhang and Kazuhito Koishida.
\newblock End-to-end text-independent speaker verification with triplet loss on
  short utterances.
\newblock In {\em Interspeech}, pages 1487--1491, 2017.

\bibitem{mo2020self}
J~Mo and L~Xu.
\newblock Self-attention networks for speaker identification with
  negative-focused triplet loss.
\newblock In {\em Journal of Physics: Conference Series}, volume 1601, page
  052004. IOP Publishing, 2020.

\bibitem{azul2013voices}
David Azul.
\newblock How do voices become gendered? a critical examination of everyday and
  medical constructions of the relationship between voice, sex, and gender
  identity.
\newblock In {\em Challenging popular myths of sex, gender and biology}, pages
  77--88. Springer, 2013.

\bibitem{mcdermott2011distinguishing}
Rose McDermott and Peter~K Hatemi.
\newblock Distinguishing sex and gender.
\newblock {\em PS: Political Science \& Politics}, 44(1):89--92, 2011.

\bibitem{ertam2019effective}
Fatih Ertam.
\newblock An effective gender recognition approach using voice data via deeper
  lstm networks.
\newblock {\em Applied Acoustics}, 156:351--358, 2019.

\bibitem{irani2011modeling}
Danesh Irani, Steve Webb, Kang Li, and Calton Pu.
\newblock Modeling unintended personal-information leakage from multiple online
  social networks.
\newblock {\em IEEE Internet Computing}, 15(3):13--19, 2011.

\bibitem{etzioni2018hippocratic}
Oren Etzioni.
\newblock A hippocratic oath for artificial intelligence practitioners.
\newblock Technical report, 2018.
\newblock Accessed 01/21/2021.

\bibitem{safavi2016fraud}
Saeid Safavi, Hock Gan, Iosif Mporas, and Reza Sotudeh.
\newblock Fraud detection in voice-based identity authentication applications
  and services.
\newblock In {\em 2016 IEEE 16th international conference on data mining
  workshops (ICDMW)}, pages 1074--1081. IEEE, 2016.

\bibitem{zimmer2010but}
Michael Zimmer.
\newblock “but the data is already public”: on the ethics of research in
  facebook.
\newblock {\em Ethics and information technology}, 12(4):313--325, 2010.

\end{thebibliography}

\end{document}